\documentclass[usenatbib]{mn2e}
\usepackage{graphicx,amssym}
\def\zsun{Z_{\odot}}
\def\<{\langle}
\def\>{\rangle}
\def\lsim{\mathrel{\hbox{\rlap{\hbox{\lower4pt\hbox{$\sim$}}}\hbox{$<$}}}}
\def\gsim{\mathrel{\hbox{\rlap{\hbox{\lower4pt\hbox{$\sim$}}}\hbox{$>$}}}}

\newcommand{\beq}{\begin{eqnarray}}
\newcommand{\eeq}{\end{eqnarray}}
\newcommand{\bc}{\begin{center}}
\newcommand{\ec}{\end{center}}
\newcommand{\Gone}{${\cal G}_1$}
\newcommand{\Gtwo}{${\cal G}_2$}

\title[The hierarchical formation of BCGs] 
      {The hierarchical formation of the brightest cluster galaxies}
\author[G. De Lucia \& J. Blaizot]
      {Gabriella De Lucia\thanks{Email: gdelucia@mpa-garching.mpg.de} and 
	J\'er\'emy Blaizot\thanks{Email: blaizot@mpa-garching.mpg.de}\\
        Max--Planck--Institut f\"ur Astrophysik, 
        Karl--Schwarzschild--Str. 1, D-85748 Garching, Germany}
\begin{document}


\pagerange{\pageref{firstpage}--\pageref{lastpage}} 
\pubyear{2006}

\maketitle
\label{firstpage}

\begin{abstract}
  We use semi-analytic techniques to study the formation and evolution of
  brightest cluster galaxies (BCGs). We show the extreme hierarchical nature of
  these objects and discuss the limitations of simple ways to capture their
  evolution. In a model where cooling flows are suppressed at late times by AGN
  activity, the stars of BCGs are formed very early (50 per cent at $z\sim 5$,
  80 per cent at $z\sim3$) and in many small galaxies. The high star formation
  rates in these high-$z$ progenitors are fuelled by rapid cooling, not by
  merger-triggered starbursts. We find that model BCGs assemble surprisingly
  late: half their final mass is typically locked-up in a single galaxy after
  $z\sim0.5$. Because most of the galaxies accreted onto BCGs have little gas
  content and red colours, late mergers do not change the apparent age of BCGs.
  It is this accumulation of a large number of old stellar populations --
  driven mainly by the merging history of the dark matter halo itself -- that
  yields the observed homogeneity of BCG properties.  In the second part of the
  paper, we discuss the evolution of BCGs to high redshifts, from both
  observational and theoretical viewpoints. We show that our model BCGs are in
  qualitative agreement with high-$z$ observations. We discuss the hierarchical
  link between high-$z$ BCGs and their local counter-parts. We show that
  high-$z$ BCGs belong to the same population as the massive end of local BCG
  progenitors, although they are not in general the same galaxies. Similarly,
  high-$z$ BCGs end-up as massive galaxies in the local Universe, although only
  a fraction of them are actually BCGs of massive clusters. 
\end{abstract}


\section{Introduction}
\label{sec:intro}

Brightest cluster galaxies (BCGs) are the very bright galaxies that inhabit the
cores of rich galaxy clusters. They are among the most luminous and most
massive galaxies in the Universe at the present epoch.  At low redshift, these
objects exhibit a small dispersion in their aperture luminosities (after
correction for systematic effects and for dependences on galaxy structure and
environment).  As a result, many early studies used BCGs as `standard candles'
for classical cosmological tests
\citep{Sandage72,GunnOke75,HoesselSchneider85,PostmanLauer95}.  BCGs lie close
to the peaks of the X--ray emission in concentrated X--ray bright clusters
\citep{JonesForman84,RheeLatour91} and their rest--frame velocities have small
offsets from those of their host clusters
\citep*{QuintanaLawrie82,ZabludoffHuchraGeller90}.  They do not appear to be
drawn from the same luminosity function as cluster ellipticals
\citep{TremaineRichstone77,Dressler78} or as bright galaxies in general
\citep{BernsteinBahvsar01}. They also exhibit different luminosity profiles
than typical cluster elliptical galaxies \citep{Oemler76,Schombert86}. All
these observations suggest that the evolution of the BCGs might be appreciably
distinct from galaxy evolution in general.

Early theoretical studies discussed the role of X--ray driven {\it cooling
  flows} \citep{Silk76,CowieBinney77,Fabian94} and {\it cannibalism} due to
dynamical friction \citep{OstrikerTremaine75,White76, MalumuthRichstone84,
  Merritt85} in the formation of these special objects.  This early work was
not however successful due to the use of a simplified cluster model.  In the
now standard cold dark matter model of structure formation, we understand that
local massive haloes (clusters) assembled rather late, through the merging of
smaller systems. In this perspective, cooling flows are indeed the main fuel
for galaxy mass-growth at high redshifts, in dense and lower mass haloes. This
source is removed only at low redshifts and in group or cluster environments,
possibly due to AGN feedback. In addition, galaxy-galaxy mergers are most
efficient within small haloes with low velocity dispersions. They are indeed
driven by dynamical friction, but it is the accretion rate of galaxies into the
proto-cluster, along with cluster growth itself, that regulates and sets the
conditions for galaxy merging.

Since these early studies, little theoretical work has focused on the formation
and evolution of BCGs, and it was only in 1998 that the subject was revisited
in a CDM hierarchical framework \citep{Dubinski98,
  Aragon-SalamancaBaughKauffmann98}.  \citet{Dubinski98} used a N--body
simulation of a cluster of galaxies and showed that merging naturally produces
a massive central galaxy with surface brightness and velocity dispersion that
resemble those of BCGs.  \citet*{Aragon-SalamancaBaughKauffmann98} analysed the
K--band Hubble diagram of BCGs up to redshift $\sim 1$ and interpreted the
apparent lack of passive evolution as evidence for recent mass accretion by the
BCGs.  Using a simple parametrisation, the authors estimated a mass growth by a
factor $4-5$, a result that they found to be consistent with predictions from
semi--analytic models of galaxy formation.  Later work
\citep{CollinsMann98,BurkeCollinsMann00,Brough02} questioned this agreement
showing that it might reflect cluster selection.  \citet{BurkeCollinsMann00}
showed that the results of \citet{Aragon-SalamancaBaughKauffmann98} were based
on a sample of low X-ray luminosity (low--L$_{\rm X}$) clusters.  For high--L$_{\rm X}$ clusters,
which better match the cluster mass selection used in the semi-analytic models,
they found substantially lower rates of mass accretion, at most a factor of
$\sim 2$ since $z\sim 1$.  In recent work, \citet{Gao04a} used numerical
simulations to study the formation of the inner cores of massive dark matter
haloes.  Their results suggest that typical central galaxies have undergone a
significant number of mergers since $z\sim1$.  Observations do provide some
examples of central galaxies in the act of merging \citep{Nipoti03}, although
not many. It is therefore still a matter of debate whether the large number of
mergers predicted in a hierarchical scenario is in agreement with observational
data.

In this paper we study the formation and evolution of BCGs using a combination
of N--body simulation and semi-analytic techniques.  In these models, central
galaxies are bound to be `special': gas that is shock heated to the virial
temperature of dark matter haloes cools radiatively and condenses only onto the
central galaxy.  An additional physical process, such as feedback from a
central AGN, must be invoked to prevent excessive growth from such cooling
flows. The haloes which BCGs inhabit originate from the gravitational collapse
of the highest (and rarest) peaks of primordial density fluctuations.  In these
overdense regions, haloes collapse earlier and merge more rapidly in comparison
to regions of the Universe with {\it average} density.  Taking advantage of the
largest simulation of the cosmic structure carried out so far -- the {\it
  Millennium Simulation} \citep{Springel05}-- we select a large number of
massive haloes at $z=0$ and study in detail how and when the stars that compose
their central galaxies formed, and how the assembly of these objects relate to
the assembly of the halo itself.  In the second part of the paper, we
investigate if the low rate of mass accretion inferred from observations is in
agreement with our model predictions.

We describe the simulation and the semi-analytic model used for this
study in Secs.~\ref{sec:sam1} and \ref{sec:sam2}.  In
Sec.~\ref{sec:case}, we present a {\it case--study} in order to define
the terminology that we will use later on. In Sec.~\ref{sec:stat} we
present statistical results for $125$ BCGs selected from the
simulation at $z=0$. In Sec.~\ref{sec:obs}, we extend that selection
to higher $z$ and we compare the luminosity evolution measured using
high--L$_{\rm X}$ clusters to that predicted by our model.  Finally,
we discuss our results and give our conclusions in
Sec.~\ref{sec:concl}.

\section{Simulation and Halo Merger Trees}
\label{sec:sam1}

In this study, we make use of the {\it Millennium
Simulation}\footnote{http://www.mpa-garching.mpg.de/galform/virgo/millennium/},
recently carried out by the Virgo
Consortium\footnote{http://www.virgo.dur.ac.uk/} and described in detail in
\citet{Springel05}.  The simulation follows $N= 2160^3$ particles of mass
$8.6\times10^{8}\,h^{-1}{\rm M}_{\odot}$ within a comoving box of size
$500\, h^{-1}$Mpc on a side.  The cosmological model is a $\Lambda$CDM
model with $\Omega_{\rm m}=0.25$, $\Omega_{\rm b}=0.045$, $h=0.73$,
$\Omega_\Lambda=0.75$, $n=1$, and $\sigma_8=0.9$, where the Hubble
constant is parameterised as $H_0 = 100\, h\, {\rm km\, s^{-1}
Mpc^{-1}}$.  These cosmological parameters are consistent with recent
determinations from the combined analysis of the 2dFGRS and first year
WMAP data \citep{SanchezEtal06}. Given its high resolution and large
volume, the Millennium Simulation allows us to follow in enough
details the formation history of a {\it representative} sample of rare
massive clusters.

During the simulation, $64$ snapshots were saved, together with group
catalogues and their embedded substructures, identified using the algorithm
{\small SUBFIND} \citep{Springel01}.  As explained in \citet{Springel05}, dark
matter haloes are identified using a standard friends--of--friends (FOF)
algorithm with a linking length of $0.2$ in units of the mean particle
separation.  The FOF group is then decomposed into a set of disjoint
substructures, each of which is identified by {\small SUBFIND} as a locally
overdense region in the density field of the background halo.  Only the
identified substructures which retain after a gravitational unbinding
procedure at least $20$ bound particles are considered to be genuine
substructures.  We note that {\small SUBFIND} classifies all the particles
inside a FOF group either as belonging to a bound substructure or as being
unbound.  The self-bound part of the FOF group itself will then also appear in
the substructure list and represents what we will refer to below as the {\it
  main halo}. This particular subhalo typically contains 90 per cent of the
mass of the FOF group \citep[e.g.][]{Springel01}.  The group catalogues were
then used to construct detailed merging history trees of all gravitationally
self-bound dark matter structures.  These merger trees form the basic input
needed by the semi-analytic model used here.

In the present paper, we use an improved scheme for tracking halo
central galaxies, which we briefly explain in the following.  In
\citet{Springel05}, the ``first progenitor'' of a given halo was
simply defined as the most massive of its progenitors. This pointer
was meant to efficiently track the main branch of a merger tree,
i.e. the branch containing the central galaxy of a FOF group. However,
this selection is sometimes ambiguous, for example when there are two
subhalos of similar mass such that their rank order in size is subject
to noise. In order to avoid such occasional failures, we have
redefined the most-massive progenitor branch by using a different
criterion to determine the ``first progenitor'' designation.  Briefly,
let $N_i$ denote the number of self-bound particles of halo $i$. Then
we define a quantity $M_i$ as
\begin{equation}
M_i = N_i +\max(M_{i_1}, M_{i_2}, \ldots, M_{i_n}), \label{eqnfirstselect}
\end{equation}
where the maximum is taken over all the progenitors $i_j$ of halo
$i$. This definition is applied recursively to each halo, over all
timesteps, and sets $M_i$ to the mass of the main branch rooted in
halo $i$. The ``first progenitor'' is then selected as the progenitor with
the largest value of $M_i$.  This definition therefore selects, for any halo,
the branch that accounts for most of the mass of the final system for the
longest period.  We argue that this provides a reasonable and very robust
definition of the ``most massive progenitor history''. Note that the scheme
adopted in \citet{Springel05} for selecting the main progenitor branch
corresponds to replacing equation (\ref{eqnfirstselect}) with $M_i=N_i$.

\section{Semi-analytics} \label{sec:sam2}

The semi-analytic model we use in the present paper is a slightly modified
version of that used in \citet{Springel05}, \citet{Croton06}, and
\citet{DeLucia06}. This model builds on the methodology introduced by
\citet{KauffmannHaehnelt00}, \citet{Springel01} and \citet*{DeLucia04b}, and
includes a model for the suppression of cooling flows by ``radio mode''
feedback from AGN \citep{Croton06}. The reader is referred to these papers for
a detailed description of the physical processes included in the model. In this
section, we simply summarize our additions and changes relative to
\citet{Croton06} (Secs. \ref{sec:bla1} and \ref{sec:bla2}), and remind the
reader with our treatment of some of the processes that are relevant for the
present work (Sec. \ref{sec:mergers}). The prescriptions and parameter values
described below are the ones that have been used to generate our standard model
accessible online \citep{LemsonVirgo06}.

\subsection{Galaxy Mergers} \label{sec:mergers}

We treat galaxy mergers as in \citet{Springel01} and \citet{Croton06}.
Substructures allow us to follow properly the motion of the galaxies sitting at
their centres until tidal truncation and stripping disrupt the subhalos at the
resolution limit of the simulation (here $1.7\times10^{10}M_{\sun}\,h^{-1}$)
\citep{Ghigna00,Kravtsov04,DeLucia04a,Gao04b}.  When this happens, we estimate
a survival time for the galaxies using their current orbit and the classical
dynamical friction formula of \citet{BinneyTremaine87}.  After this time, the
galaxy is assumed to merge onto the central galaxy of its own halo.  This is
usually the main halo of the FOF group but can be a proper substructure within
it.  In this paper, we have opted to increase the merging time estimates by a
factor of $2$ compared to the dynamical friction formula used by
\citet{Croton06}.  This change is not essential for the success of the model
but it slightly improves the fit of the bright end of the luminosity function.
The adjustment is somewhat {\it ad-hoc} but seems justified given that the
terms that enter the estimation of the dynamical friction times carry large
uncertainties (most notably the Coulomb logarithm and the orbital
distribution), and that the correct pre-factor has not been calibrated yet
with detailed numerical studies at the resolution we have here.  In addition,
there are some direct indications that there are problems with these formulas:
\citet{Springel01} noted that when haloes of comparable mass merge, the
inferred merging times are typically shorter than the time measured using high
resolution numerical simulations.

A galaxy merger is accompained by a starburst modelled using the ``collisional
starburst'' prescription introduced by \citet{SomervilleEtal01}. Say two
galaxies \Gone{} and \Gtwo{} of masses $m_1 > m_2$ merge together. We assume
that all the gas from \Gone{} and \Gtwo{} is gathered in the disk component of
the remnant galaxy ${\cal G}$, while all the stars of \Gtwo{} are added to the
bulge stars of \Gone{} into the buge component of ${\cal G}$.  A fraction of
the gas in ${\cal G}$ is then converted instantaneously into stars, the
fraction depending on the baryonic mass ratio of the two merging galaxies:
\begin{eqnarray}
\nonumber
m_{\rm star}^{\rm new} &=& 0.56 \times \left( \frac{m_2}{m_1} \right)^{0.7}
\times m_{\rm gas} 
\end{eqnarray}
The numerical parameters in the equation above provide a good fit to the
numerical results of \citet{CoxEtal2004}.  We note that the above formulation
implies that in the case of an equal mass merger ($m_1 \sim m_2$), about 40\%
of the gas is turned into stars\footnote{The fraction of gas that is turned
into stars during a major merger saturates at 40\% because of the competition
of feedback, as explained in \citet{Croton06}.}, while in the case where $m_2
<< m_1$, no star formation is triggered by the interaction. In case of a major
merger, which occurs when $m_2/m_1 > 0.3$, we further assume that the final
disc is completely disrupted and all the stars of galaxy ${\cal G}$ are put in
a bulge component.

The formation of new stars is accompanied by the ejection of a fraction
of the remaining cold gas, which is modelled using the same prescription
as adopted in \citet[][see their Sec.~3.6]{Croton06}.  Note that this
feedback model is different than that employed in
\citet{DeLucia06} and results in more efficient ejections especially
for massive galaxies.  For the purposes of this work, these two schemes
do not produce significantly different results, except for
Figs.~\ref{fig:evol1} and \ref{fig:evol2}, which we discuss in
Sec.~\ref{sec:obsevol}.

\subsection{Stellar Populations} \label{sec:bla1}

The photometric properties of our model galaxies are computed employing the
stellar population synthesis model from \citet{BruzualCharlot03} and using the
method described in \citet{DeLucia04b}. Contrary to \citet{Croton06}, we adopt
here the initial mass function (IMF) from \citet{Chabrier03}, together with the
Padova 1994 evolutionary tracks.  The Chabrier IMF can be represented by a
power-law with index 1.3 from 1 to 100 M$_\odot$, and a lognormal distribution
from 0.1 to 1 M$_\odot$, centered on 0.08 M$_\odot$ and with dispersion 0.69
\citep[see][]{BruzualCharlot03}.  We note that the spectral properties obtained
using this IMF are very similar to those obtained using the Kroupa (2001) IMF.
In particular, a Chabrier IMF yields a larger fraction of massive stars per
unit mass than Salpeter IMF, and hence a mass-to-light ratio typically about
1.8 times lower in the $V$ band \citep[see Fig. 4 of][]{BruzualCharlot03}.  The
Chabrier IMF is physically motivated and provides a better fit to counts of
low-mass stars and brown dwarfs in the Galactic disc \citep{Chabrier03}.

Because of the substantial differences between the Salpeter and Chabrier IMFs,
we had to re-adjust some of the model parameters in order to obtain a good
agreements with observational results in the local Universe.  With respect to
Table 1 in \citet{Croton06}, we had to increase $\kappa_{\rm{AGN}}$ (the
quiescent hot gas BH accretion rate) from $6\times 10^{-6}\,{\rm M_{\odot}
  yr^{-1}}$ to $7.5\times 10^{-6}\,{\rm M_{\odot} yr^{-1}}$ and decrease
$\alpha_{\rm{SF}}$ (the star formation efficiency) from $0.07$ to $0.03$.
Finally, for consistency with the use of a Chabrier IMF, we adopt an
instantaneous recycled fraction $R = 0.43$.

\subsection{Dust attenuation} \label{sec:bla2}

In this work we use a new parametrisation for dust attenuation, which
combines those developed in \citet*{DevriendtGuiderdoniSadat99} -- for
a homogeneous inter-stellar medium (ISM) component -- and \citet{CF00}
-- for molecular clouds around newly formed stars.

We assume that the mean perpendicular optical depth
of a galactic disk at wavelength $\lambda$ is:
\begin{displaymath}
\tau_\lambda^{\mathrm{z}}  = \left( \frac{A_\lambda}{A_{\mathrm{V}}}
\right)_{\zsun} \left( \frac{Z_\mathrm{g}}{\zsun} \right)^s  \left(
 \frac{ \left<{N_\mathrm{H}}\right>}{2.1\  10^{21} \mathrm{atoms \, cm^{-2}}} \right)
\end{displaymath}
where the mean H column density is given by: 
\begin{displaymath}
\left<{N_\mathrm{H}}\right> = \frac{M_{\rm cold}}{1.4 m_p \pi (a\,r_{1/2})^2}\mathrm{atoms \, cm^{-2}}.
\end{displaymath}
In the previous equation, $r_{1/2}$ is the half-mass radius of the disk and
$a=1.68$ is such that the column density represents the mass-weighted average
column density of the disk, which is assumed to be exponential. As in
\citet{Hatton03} -- see \citet{GuiderdoniRoccaVolmerange87} for details -- the
extinction curve depends on the gas metallicity $Z_g$ and is based on an
interpolation between the Solar neighbourhood and the Large and Small
Magellanic Clouds ($s = 1.35$ for $\lambda < 2000\AA$ and $s = 1.6$ for
$\lambda > 2000\AA$).  The adopted extinction curve for solar metallicity is
that of \citet*{MMP83}.  Finaly, we assign a random inclination to each galaxy
and apply the dust correction to its disc component using a `slab' geometry
\citep[see][]{DevriendtGuiderdoniSadat99}.

In addition to this extinction from a diffuse ISM component, we have
also implemented a simple model to take into account the attenuation
of young stars within their birth clouds, based on \citet{CF00}.
Stars younger than the finite lifetime of stellar birth clouds (which
we assume to be equal to $10^7\,{\rm yr}$) are subject to a
differential attenuation with mean perpendicular optical depth:
\begin{displaymath}
  \tau_\lambda^{\mathrm{BC}} = \tau_{\mathrm{V}}^{\mathrm{BC}} \left
  (\frac{\lambda}{5500\AA}\right)^{-0.7}
\end{displaymath} 
and 
\begin{displaymath}
\tau_{\mathrm{V}}^{\mathrm{BC}} = \tau_\lambda^{\mathrm{z}} \times \left(\frac{1}{\mu} -1\right),
\end{displaymath}
where $\mu$ is drawn randomly from a Gaussian distribution with centre $0.3$
and width $0.2$, truncated at $0.1$ and $1$ \citep[see][]{KCBF04}.

\section{A case study}
\label{sec:case}

In the framework of a hierarchical scenario, the history of a galaxy is fully
described by its complete merger tree.  Whereas in the monolithic approximation
the history of a galaxy can be described by a set a functions of time,
hierarchical histories are difficult to summarise in a simple form, because
even the {\it identity} of a galaxy is sometimes ill-defined.  In hierarchical
models, ``a galaxy'' should not be viewed as a single object evolving in time,
but rather as the ensemble of its {\it progenitors} at each given time.  In
this section, we focus on a single model BCG and describe the formation and
assembly history of its stars in detail.  This allows us to define and
illustrate the behaviour of quantities that capture different aspects of 
BCG evolution.  We will use these quantities later (in Sec. \ref{sec:stat}) to
describe BCG formation in a statistical fashion.

\subsection{Trees and branches -- definitions}

\begin{figure*}
\bc
\hspace{-1.4cm}
\resizebox{18cm}{!}{\includegraphics[angle=90]{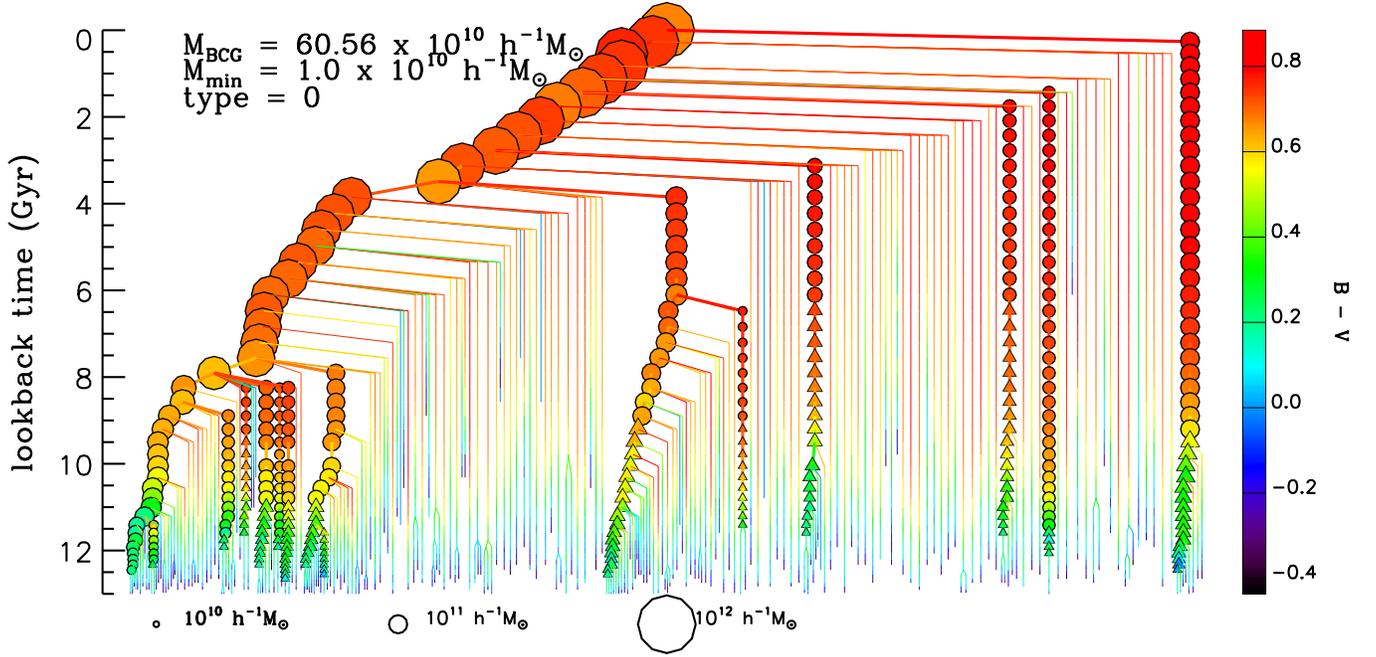}}\\%
\caption{BCG merger tree. Symbols are colour--coded as a function of
  B~-~V colour and their area scales with the stellar mass.  Only progenitors
  more massive than $10^{10}\,{\rm M}_{\odot}\,h^{-1}$ are shown with symbols.
  Circles are used for galaxies that reside in the FOF group inhabited by the
  main branch. Triangles show galaxies that have not yet joined this FOF
  group.}
\label{fig:bcg_tree}
\ec
\end{figure*}

\begin{figure*}
\bc
\hspace{-1.4cm}
\resizebox{18cm}{!}{\includegraphics[angle=90]{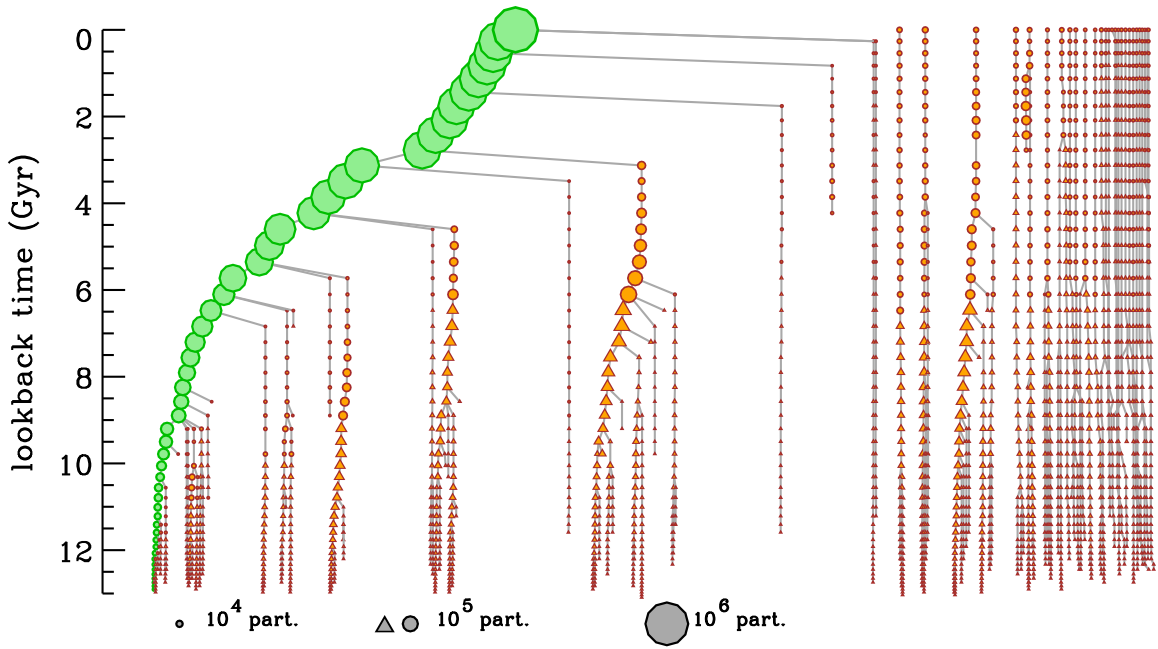}}\\%
\caption{Merger tree of the FOF group in which the BCG sits at
  redshift zero. Only the trees of subhalos with more than $500$ particles at
  $z=0$ are shown.  Their progenitors are shown down to a 100 particle limit.
  Symbol coding is the same as in Fig. \ref{fig:bcg_tree}. The left-most tree
  is that of the main subhalo of the FOF, while the trees on the right
  correspond to other substructures identified in the FOF group at $z=0$.  In
  green, we mark the subhalo that contains the main branch of the BCG.}
\label{fig:fof_tree}
\ec
\end{figure*}

Fig.~\ref{fig:bcg_tree} shows the full merger tree of the central galaxy of a
dark matter halo of mass ${\rm M}_{200} = 8.9\times 10^{14}\,{\rm M}_{\odot}$
at $z=0$. The BCG itself lies at the top of the plot (at $z=0$), and all its
progenitors (and their histories) are plotted downwards going back in time
recursively. Galaxies with stellar mass larger (resp. smaller) than $10^{10}$
$h^{-1}$M$_{\odot}$ are shown as symbols (resp. lines), and are colour-coded as
a function of their rest-frame B~-~V colour.  The left-most branch in
Fig.~\ref{fig:bcg_tree} represents what we will hereafter refer to as the {\it
  main branch}.  This particular branch is obtained by connecting the galaxy at
each time-step to the progenitor with the largest stellar mass (the {\it main
  progenitor}) at the immediately preceding time-step.  It is tempting to
consider this branch as {\it the BCG} itself, and the objects that merge onto
it as other {\it progenitors} of the BCG.  A clear distinction between ``main''
and ``other'' progenitors is indeed often quite appropriate - typically for
late type galaxies - and it is satisfactory as long as galaxies merging onto
the main branch have stellar masses that are much smaller than the current mass
of the main progenitor.  In this case, the evolution of the BCG can be
characterised by a series of accretion events (of much smaller objects) that do
not introduce major changes in the stellar population or identity of the BCG
itself.

This simple summary of a merger tree is however not sufficient to
describe histories as complex as those of BCGs \citep[see
also][]{NeisteinEtal06}. Fig.~\ref{fig:bcg_tree} shows that in our
case-study, the main branch captures the evolution of the BCG itself
for the last $\sim8$~Gyr. Before that time, the main progenitor of the
BCG is only marginally more massive than progenitors in other
branches.  At this point, choosing one branch or another becomes
arbitrary and a single branch is certainly a poor proxy for describing
the evolution of the BCG and of its stellar population.  In the
following, we will simply refer to the main progenitor of the BCG at
any given time as the main branch without implying that it necessarily
contains most of the stars.

Galaxies that merge onto the main branch must first be accreted onto the same
halo, and it is therefore interesting to establish the connection between the
galaxy and the halo merger trees.  Fig.~\ref{fig:fof_tree} shows the full tree
of the FOF-group containing our case-study BCG. The branch highlighted in green
is the branch containing the main branch of the BCG. The right-most branches
are merger trees of secondary substructures (only those with more than 500
particles) present in the FOF-group. These substructures have not yet dissolved
into the main halo, and their galaxies can thus not contribute to the merger
tree of the BCG. In Figs.~\ref{fig:bcg_tree} and \ref{fig:fof_tree} circles
mark objects (galaxies or haloes) that belong to the same FOF group as the main
branch of the BCG, while triangles mark objects that have not yet joined the
FOF group. Typically, when a halo is accreted onto a bigger system (i.e. joins
the same FOF group), it loses mass efficiently due to tidal stripping
\citep{Ghigna00,Kravtsov04,DeLucia04a,Gao04b}.  A nice example of this process
is shown by the halo branch located roughly at the centre of
Fig.~\ref{fig:fof_tree}. It is only when the subhalo dissolves that its
galaxies become part of the main halo of the FOF group and are then allowed to
merge with the central galaxy on a dynamical friction timescale.

Given the complexity of the merger history shown in
Fig.~\ref{fig:bcg_tree}, it is helpful to define several times that
mark important phases in the evolution of a BCG. We call {\it identity
time} ($t_{\rm{id}}$) the cosmic time when the BCG acquires its final
identity. We define $t_{\rm{id}}$ as the time when the last major
merger on the main branch occured, i.e. when the most massive galaxy
which merges on the main branch is more massive than a third of the
mass of the main progenitor. Before $t_{\rm{id}}$, the BCG does not
exist as one single object, but as several progenitors of comparable
masses.  Our definition of the identity time can be extended to
account for multiple simultaneous mergers.  We thus define the {\it
extended identity time} ($\tilde{t}_{\rm{id}}$) as the latest (cosmic) time
when the sum of the masses of the progenitors merging on the main
branch was larger than a third of the mass of the main progenitor.  By
definition, $t_{\rm{id}} \leq \tilde{t}_{\rm{id}}$, although they are
equal in most cases (see Fig.  \ref{fig:stat_times} in
Sec.~\ref{sec:stat}).

Early theoretical work discussed the difference between
formation and assembly times for elliptical galaxies in a hierarchical context
\citep*{BaughColeFrenk96,Kauffmann96}, although this difference has
been quantified only recently \citep{DeLucia06}. Following this work,
we call {\it assembly time} ($t_a$) the time when the main progenitor
contains half the final stellar mass of the BCG.  As discussed above,
although the main branch may capture the identity of a galaxy for some
time, it is not suited to describe the complete evolution of the BCG.
The stellar population of the BCG, in particular, can be fully
described only by taking into consideration the whole tree, because a
large fraction of stars actually form in secondary branches. It is
therefore useful to define a more classical {\it formation time}
($t_f$) as the time when the total mass of stars formed reaches half
the final mass of the BCG.  By ``total mass'' we mean, at each
cosmic time, the sum of the stellar masses of {\it all} the
progenitors present at that given time, i.e. the projection of
Fig.~\ref{fig:bcg_tree} onto the vertical time axis.

\subsection{Mass build-up of the BCG}

Fig.~\ref{fig:bcg_tree} shows that the stars that end up in the BCG today,
start forming at very high redshifts.  Rapid cooling in the early phases of the
cluster collapse lead to the formation of a massive central galaxy of stellar
mass $\sim 10^{11}\,{\rm M}_{\sun}\,h^{-1}$ about $11$~Gyr ago ($z\sim 2.5$).
A number of accretions of massive satellites increases its mass to slightly
less than half its present value at redshift about $1$ and then the BCG
continues growing by accretion of satellites.  Fig.~\ref{fig:bcg_tree} also
clearly shows that the satellites accreted below redshift $\sim 1$ are red and
much less massive than the main branch.

\begin{figure}
\bc
\hspace{-0.6cm}
\resizebox{9cm}{!}{\includegraphics[]{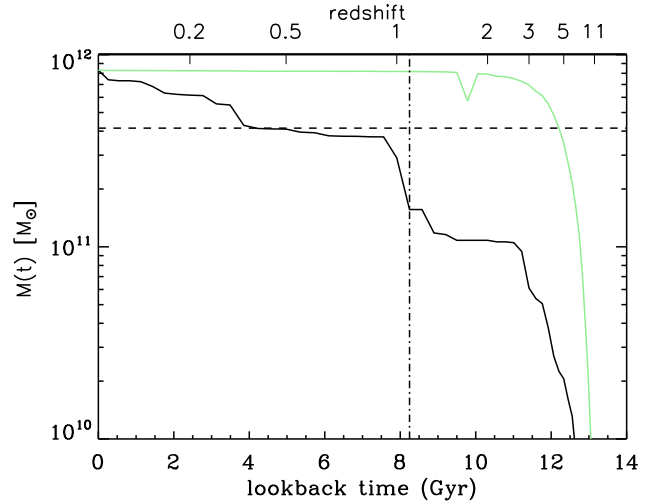}}
\caption{Formation and assembly histories of our case-study BCG. The
  black (resp. green) line shows the stellar mass in the main branch (resp. the
  mass summed over all progenitors) of our case-study BCG versus time. The
  horizontal dashed line corresponds to half the present mass of the BCG. The
  vertical lines mark the identity and extended identity times (see text for
  details).}
\label{fig:mbu1}
\ec
\end{figure}

In Fig.~\ref{fig:mbu1} we show the `formation' and `assembly' histories of the
stars that end up in the BCG of our case-study cluster. The black line in
Fig.~\ref{fig:mbu1} shows the stellar mass of the main branch.  The green line
shows the sum of the stellar masses in all progenitors at each time.  The
horizontal dashed line corresponds to half the present mass of the BCG. This
figure clearly shows that while the main progenitor reaches half its final mass
only at redshift $z_a \sim 0.5$, half of the stars are already formed at
redshift $z_f \sim 5$.  The identity time and the extended identity time are
equal in this particular case and are indicated by the vertical line at $z \sim
1.2$.

It is now interesting to investigate in more detail how the BCG gained its mass
and how the stars that end up in it were formed in the first place. Was star
formation mainly triggered by starbursts in an early and dense environment? Or
was star formation mostly quiescent, fuelled by rapid cooling?

\begin{figure}
\bc
\hspace{-0.6cm}
\resizebox{9cm}{!}{\includegraphics[]{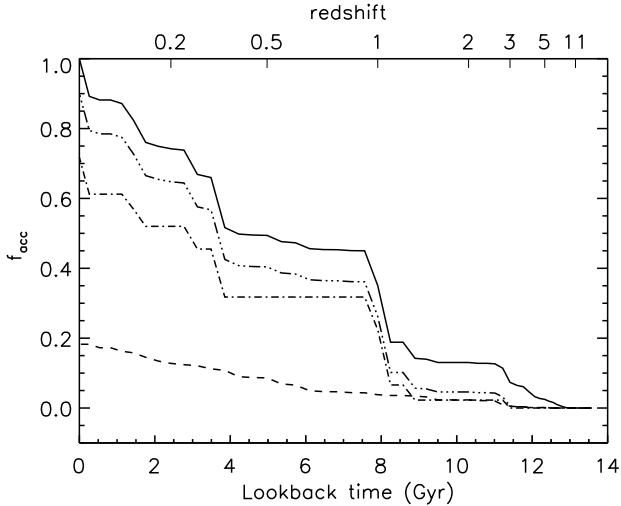}}
\caption{Accretion history of our case-study BCG~: fraction of the stellar mass
that was brought by mergers onto the main branch. The solid line is the mass of
the main branch. The 3-dot-dashed (resp. dot-dashed, dashed) line is the mass
accreted by all (resp. more massive than $10^{10}$~$h^{-1}$M$_\odot$,
less massive than $10^{10}$~$h^{-1}$M$_\odot$) galaxies merging on the
main branch.}
\label{fig:mbu2}
\ec
\end{figure}

Fig.~\ref{fig:mbu2} shows the mass in the main branch as a function of
time (solid line), and the fraction of it that was gained through
accretion of galaxies (dash-3-dot curve). The dashed
(resp. dot-dashed) curve indicates the contribution to accretion by
galaxies with stellar masses lower (resp. larger) than
$10^{10}$~$h^{-1}$M$_\odot$ (the same cut as in
Fig. \ref{fig:bcg_tree}).  Fig.~\ref{fig:mbu2} shows that most of the
stars of this BCG were actually not formed in the main branch, but
were instead accreted steadily over time.  It is also interesting to
note that, although the BCG tree has many branches, only a small
fraction of these contribute significantly to the build-up of its
mass: $\sim70$ per cent of the mass comes from accretion of $12$
galaxies more massive than $10^{10}$~$h^{-1}$M$_\odot$.  Note also
that the main progenitor becomes rapidly massive enough that most
galaxies falling onto the main branch give rise only to minor merger
events. Fig.~\ref{fig:mbu2} shows that since $z\sim4$ the mass growth
of the main branch is only due to accretion: AGN activity has switched
off cooling and hence star formation in the main branch.  The stars
that compose the BCG today formed in separate entities, mostly as a
result of quiescent star formation: less than $10$ per cent of the
stars were formed during starbursts \citep[in agreement with previous
results by][]{Aragon-SalamancaBaughKauffmann98}. Note that this
quiescent-to-bursty star formation ratio depends on the star
formation prescription that we use for
discs. The results of \citet{SomervilleEtal01} and
\citet{BaughEtal05} suggest that our model predicts an upper boundary
of this ratio.

\begin{figure}
\bc
\hspace{-0.6cm}
\resizebox{9cm}{!}{\includegraphics[]{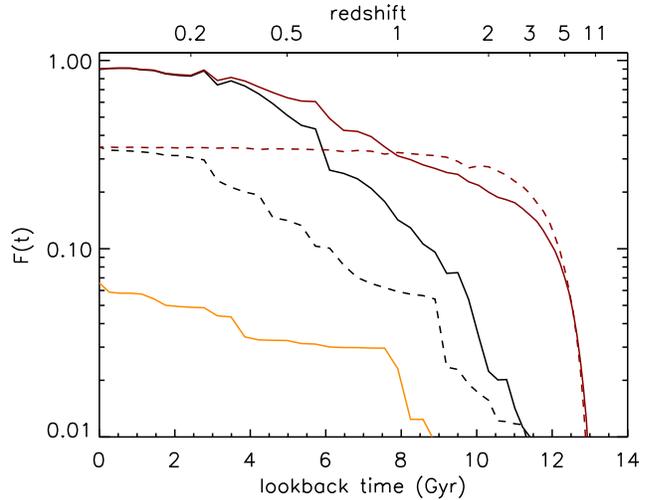}}
\caption{Evolution of the stellar content (dashed lines) and dark matter
  content (solid lines) of the subhalo in which our case-study BCG sits at
  $z=0$.  Black and red lines are obtained by following the main progenitor
  only and all progenitors at any given redshift respectively. Dark matter
  masses are normalised to the dark matter mass of the FOF group at redshift
  zero, while stellar masses are normalised to the total stellar content of the
  FOF group at redshift zero.  The orange line shows the stellar mass in the
  main branch.}
\label{fig:mbu4_5}
\ec
\end{figure}

In a hierarchical framework, the formation of a BCG naturally relates to the
formation of the cluster itself.  Fig.~\ref{fig:mbu4_5} shows how the stellar
mass in the main branch (in orange) compares to the evolution of the dark
matter (solid black line) and stellar content (dashed black line) of the halo
in which the main progenitor of the BCG resides at each time.  The red lines
show instead the evolution in dark matter (solid) and stellar content (dashed)
of all progenitors of the halo in which the galaxy sits today.  Dark matter
masses are normalised to the dark matter mass of the FOF group at redshift
zero, while stellar masses are normalised to the total stellar content of the
FOF group at redshift zero.  The figure shows that the assembly history of the
BCG follows the evolution of the dark matter and stellar content of its own
halo with a delay due to dynamical friction.  The parent halo of the BCG
represents only $90$ per cent of the total mass of its parent FOF group, and
its total stellar content accounts for about $35$ per cent of the stellar
content of the FOF group. Note that although dark matter substructures only
represent $\sim 10$ per cent of the mass of the FOF group, they contain about
$65$ per cent of its stellar content.  Interestingly, the total stellar content
of all the progenitors of the halo increases very rapidly and stays constant
after redshift $\sim 1$.  This indicates that many substructures accreted below
redshift $\sim 1$ survive as independent entities to the present day so that
their galaxies do not contribute to the stellar content of the halo.

\section{Statistical results}
\label{sec:stat}

In the previous section we have analysed in detail the formation and assembly
history of one BCG chosen as a case study, and we have related these histories
to the history of the parent halo itself.  In this section, we extend this
analysis to a statistical sample of BCGs selected from our simulation and study
the dispersion in the quantities defined above.  This is of particular interest
because observed BCGs exhibit a remarkably small dispersion in their
luminosities (see Sec.~\ref{sec:intro}) and stellar population at low redshift
\citep[e.g.][]{McNamaraOConnell89, JamesMobasher00}.  This is commonly
interpreted as a clear indication that these objects had similar formation
histories.  In order to address this point, we have selected the 125 haloes
more massive than $7\times 10^{14}$ M$_\odot$ at $z=0$ from the Millennium
Simulation and analysed their formation and assembly histories, as in the
previous section.  The BCGs of these objects have mean absolute magnitude ${\rm
  M}_{\rm K} = -26.58$ and a dispersion of $0.20$~mag.  These values appear to
be in nice agreement with observational results \citep[see for example Fig.~10
in][]{CollinsMann98}.  This agreement represents a major success of the
underlying galaxy formation model. 

\begin{figure}
\bc
\hspace{-0.63cm}
\resizebox{9cm}{!}{\includegraphics[]{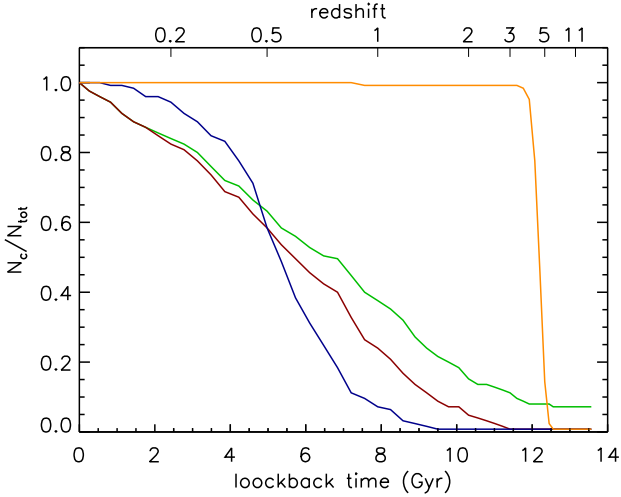}}
\caption{Fraction of clusters with identity (green), extended identity (red),
  assembly (blue) and formation (orange) lookback times larger than
  that shown on the x-axis.}
\label{fig:stat_times}
\ec
\end{figure}

In Fig.~\ref{fig:stat_times} we show the fraction of clusters with
identity (green), extended identity (red), assembly (blue), and
formation (orange) lookback times larger than that shown on the
x-axis.  The green line in Fig.~\ref{fig:stat_times} shows that for
only about $40$ per cent of our BCGs was the last major merger before
redshift $1$, and for $\sim 65$ per cent of them it occurred before
redshift $\sim 0.5$.  The numbers are slightly lower if one considers,
instead of the time corresponding to the last major merger, the time
when the sum of the masses of the accreted objects is equal to one
third the mass in the main branch at the time of accretion (the
extended identity time). The identity of the main progenitor of a
redshift zero BCG is thus typically ill-defined before $z\sim
0.7$. The assembly times measured for our BCGs are low~: half of them
assembled after $z\sim 0.5$, and only 10 per cent of our BCGs
assembled before $z\sim 1$.  In contrast, for almost $100$ per cent of
our BCGs, half of the stars were already formed at redshift larger
than $4$, which indicates that all our BCGs have uniformly old stellar
populations.

\begin{figure}
\bc
\hspace{-0.6cm}
\resizebox{9.cm}{!}{\includegraphics[]{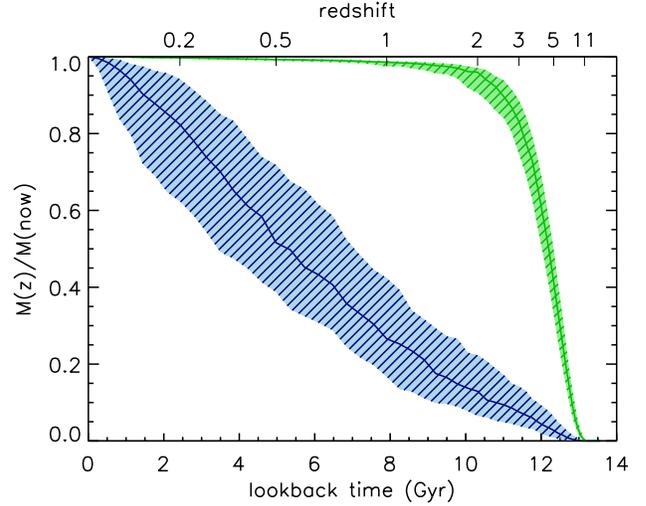}}
\caption{Assembly (blue) and formation (green) histories of our sample of BCGs
  selected at redshift $0$ (as in Fig.~\ref{fig:mbu1}). Thick lines show the
  median of the distributions, while the dashed regions show the $15$th to
  $85$th percentile range.}
\label{fig:stat_ass_form}
\ec
\end{figure}

Thick lines in Fig.~\ref{fig:stat_ass_form} show the median assembly (blue) and
formation (green) histories of our sample of redshift zero BCGs (as in
Fig.~\ref{fig:mbu1}).  The dashed regions in this figure indicate the $15$th
and $85$th percentiles of the distributions.  Interestingly, there is a very
small scatter in the formation histories of the BCGs: for essentially all the
objects in our sample, $50$ per cent of the stars are already formed at
redshift $5$ (as noticed from Fig.~\ref{fig:stat_times}).  The assembly
histories exhibit a much larger scatter with the fraction of mass in the main
progenitor varying between $15$ and $40$ per cent at redshift $1$, and between
$40$ and $70$ per cent at redshift $0.5$.  In Sec.~\ref{sec:case}, we have seen
that the mass of the BCG mainly increases through accretion of satellites,
Fig.~\ref{fig:stat_ass_form} shows that this is the case for the whole sample.
The difference between formation time and assembly time that we find here is
higher than that found by \citet{DeLucia06} for ellipticals, as expected from
the ``extreme'' nature of BCGs.  We note also that the we find a median mass
growth of a factor $\sim 3$ from $z\sim 1$ to $z=0$.  This is in apparent
contradiction with observational data mentioned in Sec.~\ref{sec:intro}.  We
will show that this is probably only an `apparent' contradiction in
Sec.~\ref{sec:obsevol}.

\begin{figure}
\bc
\hspace{-0.6cm}
\resizebox{9.cm}{!}{\includegraphics[]{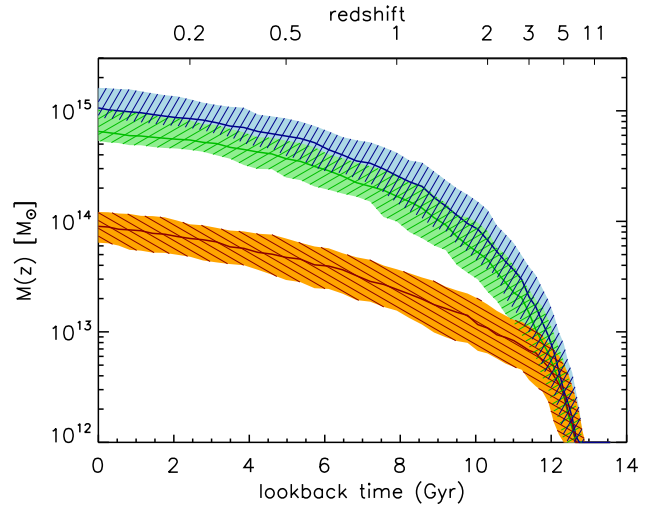}}
\caption{Evolution of the dark matter mass of the FOF group containing the main
  progenitor of the BCG at each time (blue), of the stellar content of this FOF
  group (green), and of the mass of the main progenitor of the BCG (orange).
  Thick lines show the median and shaded regions show the $15$th to $85$th
  percentile range.  Note that we have multiplied the stellar mass of the FOF
  group by a factor of $50$ and the mass of the main progenitor of the BCG by a
  factor of $100$.}
\label{fig:stat_halo_masses}
\ec
\end{figure}
 
Fig.~\ref{fig:stat_halo_masses} shows how the evolution of the mass in the main
progenitor (orange) compares to the evolution in the stellar (green) and dark
matter (blue) content of the FOF group in which the main progenitor resides at
each lookback time.  Thick lines show the median and shaded regions show the
$15$th and the $85$th percentiles of the distributions.  Note that we have
multiplied the stellar mass of the FOF group by a factor of $50$ and the mass
of the main progenitor of the BCG by a factor $100$ for display purposes.  The
figure shows that the stellar content of the FOF group in which the BCG sits,
represents about $1.7$ per cent of its dark matter mass, at all redshifts.  In
turn, the mass in the main branch is $\sim 100$ per cent of the stellar content
of the FOF group at high $z$, and then decreases to reach a plateau at $z\sim
1$, with a mass fraction of $\sim 6$ per cent.

\begin{figure}
\bc
\hspace{-0.6cm}
\resizebox{9.cm}{!}{\includegraphics[]{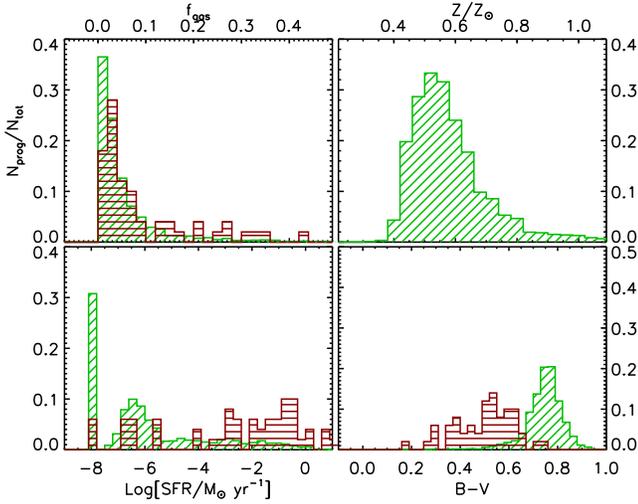}}
\caption{Distribution of objects accreted onto the main branch as a function
  of their gas fraction (panel a), star formation rate (panel b), stellar
  metallicity (panel c), and B~-~V colour (panel d).  Green histograms refer to
  all progenitors, independently of their accretion redshifts.  Red histograms
  refer to objects accreted at redshift larger than $3$.}
\label{fig:stat_progs}
\ec
\end{figure}

The example in Fig.~\ref{fig:bcg_tree} shows that most of the progenitors
accreted onto the main branch, are already red at the time of accretion.  We
now want to quantify the distribution of the objects accreted onto the main
branch in terms of different physical properties.  In Fig.~\ref{fig:stat_progs}
we show the distribution of these progenitors as a function of their gas
fraction (panel a), star formation rate (panel b), stellar metallicity (panel
c), and B~-~V colour (panel d).  Note that we are using here only progenitors
more massive than $10^{10}$~$h^{-1}$M$_\odot$, as these are the ones that
contribute most significantly to the mass build-up of the BCG (see
Fig.~\ref{fig:mbu2}).  The green histograms in all panels refer to all the
objects accreted, independently of their accretion redshifts.  The top left
panel indicates that the great majority of these progenitors have very low gas
fractions.  Only objects accreted at very high redshift have some residual gas,
as indicated by the red histograms; these refer to progenitors accreted at
redshift larger than $3$.  As a consequence, only objects accreted at very high
redshift have some residual star formation (panel b) and relatively blue
colours (panel d).  Interestingly, the metallicity distribution of progenitors
does not show any significant evolution with redshift. It peaks at $Z\sim 0.5
Z_{\sun}$ and is quite broad.

\begin{figure}
\bc
\hspace{-0.64cm}
\resizebox{9cm}{!}{\includegraphics[]{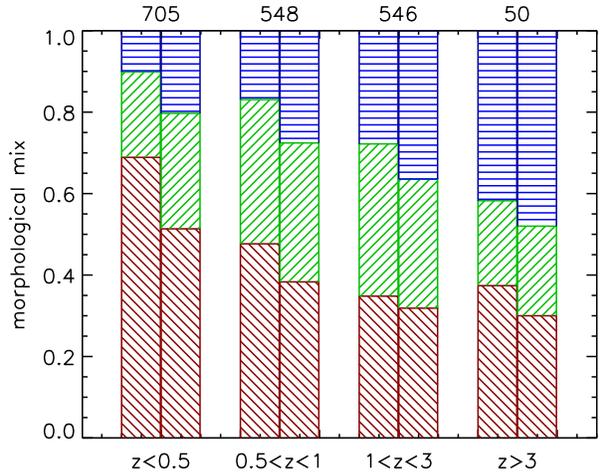}}
\caption{Morphological mix of objects accreted onto the main branch in
  different redshift bins.  The red indicates early-type morphologies, the
  green is for S0 and the blue is for late types (see text for details). For
  each redshift bin, the left-most histogram refers to a normalisation in mass,
  while the right-most histogram refers to a normalisation in number.}
\label{fig:stat_morph}
\ec
\end{figure}

Fig.~\ref{fig:stat_morph} shows the morphological mix of the objects accreted
onto the main branch in different redshift bins.  We have determined the
morphology of our model galaxies by using the B-band bulge-to-disc ratio
together with the observational relation by \citet{SimienDeVaucouleurs86}
between this quantity and the galaxy morphological type.  For the purposes of
this analysis, we classify as early types all galaxies with $\Delta M < 0.4$
($\Delta M= M_{\rm bulge} - M_{\rm total}$), as late types all galaxies with
$\Delta M > 1.56$, and as lenticulars (S0) all galaxies with intermediate value
of $\Delta M$.  In Fig.~\ref{fig:stat_morph}, the red indicates early-type
morphologies, the green is for S0 and the blue is for late types. For each
redshift bin, the left-most histogram refers to normalisation by mass, while
the right-most histogram refers to normalisation by the number of objects.  The
number of objects accreted in each redshift bin is indicated on the top of the
histograms.  The figure clearly shows that, with the exception of the highest
redshift bin considered, the majority of the objects accreted onto the main
branch, already have an early-type morphology.

\begin{figure*}
\bc
\hspace{-1.4cm}
\resizebox{18cm}{!}{\includegraphics[angle=90]{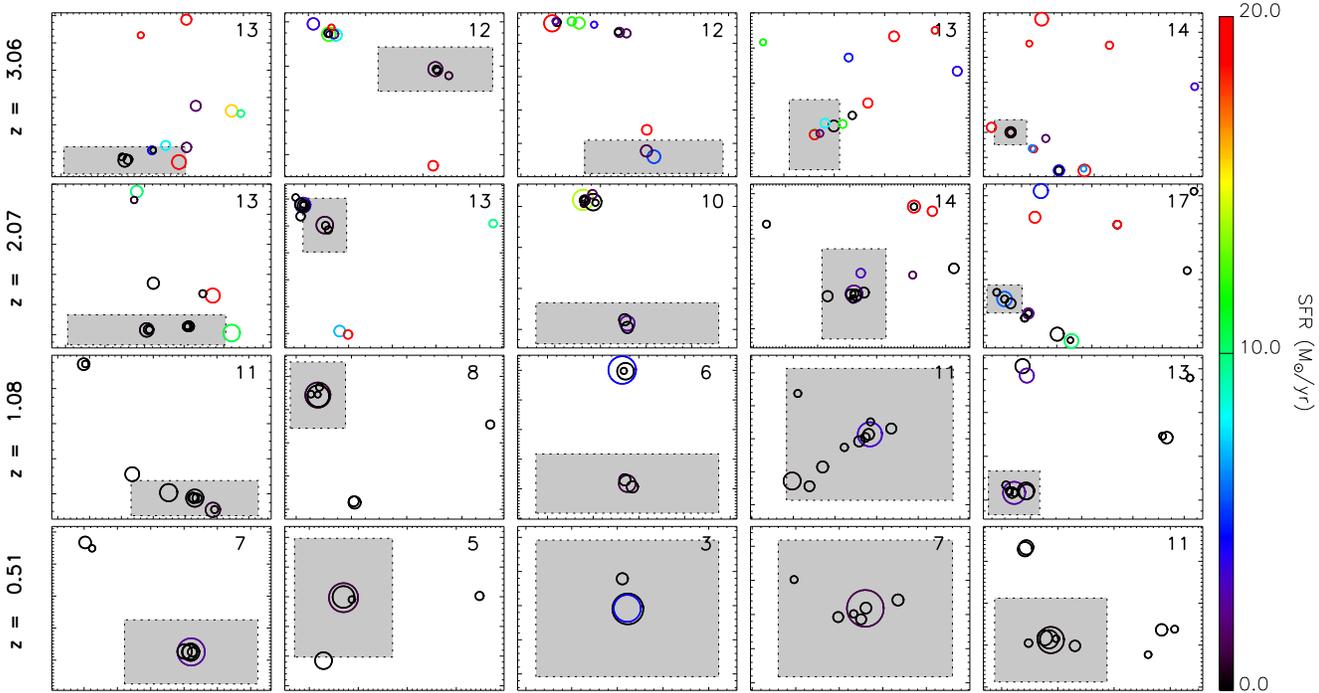}}\\%
\caption{Spatial distribution of the progenitors of five model BCGs at
  redshifts $\sim$ 0.5, 1, 2 and 3. Progenitors more massive than
  $10^{10}h^{-1}M_\odot$ are shown using symbols colour-coded with star
  formation rate and with size scaling with stellar mass. The number in the
  upper-right corner of each panel indicates the number of such progenitors.
  The gray area marks a $2\times2$~Mpc$^2$ comoving region centred on the main
  progenitor of the redshift zero BCG.}
\label{fig:progs_pos}
\ec
\end{figure*}

We saw previously that a BCG is {\it not} well described by the monolithic
approximation. Instead, at $z>0$, a BCG breaks into an ensemble of distinct
galaxies, and it is {\it all} of them at the same time. A question relevant
both to the theoretical and observational sides is then: what is the spatial
extent of the progenitor set as a function of time? In other words, how far do
the progenitors of the BCG extend?  Fig. \ref{fig:progs_pos} shows the spatial
distribution of progenitors of five model BCGs at different redshifts. Only
progenitors more massive than $10^{10}$~$h^{-1}$M$_\odot$ are shown, with
symbols that scale with mass as in Fig. \ref{fig:bcg_tree}. The symbols are
colour-coded by the star formation rate averaged over a time-step ($\sim 200$
Myr).  The number of progenitors more massive than $10^{10}h^{-1}M_\odot$ is
indicated in the upper-right corner of each panel.  The gray area marks a
$2\times2$~Mpc$^2$ comoving region and is centred on the main progenitor of the
BCG.  The typical extent of the progenitors goes from $\sim 10 h^{-1}$Mpc
(comoving) at $z=3$ down to about $\sim 1 h^{-1}$Mpc at $z=0.5$.  Several
panels (e.g. middle panels down to $z\sim 1$) show that the main progenitor of
the BCG is not necessarily the most massive progenitor at that redshift.

\begin{figure}
\bc
\hspace{-0.5cm}
\resizebox{8.7cm}{!}{\includegraphics[]{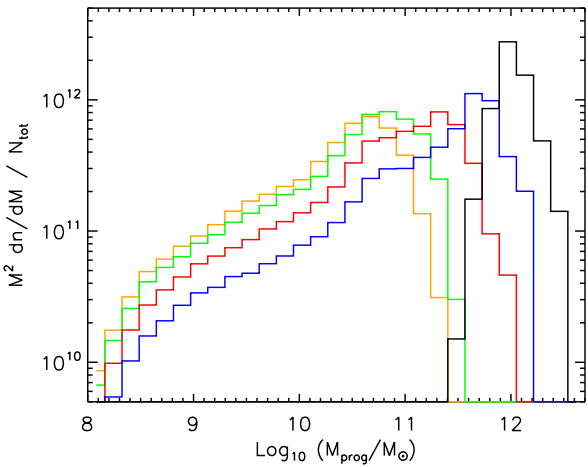}}
\caption{Stellar mass distribution for the progenitors of local BCGs at
  redshifts 0.5 (blue), 1 (red), 2 (green), and 3 (orange).  The black
  histogram shows the stellar mass distribution of local BCGs.}
\label{fig:stat_progmasses}
\ec
\end{figure}

Fig.~\ref{fig:stat_progmasses} shows the stellar mass distribution of
progenitors of redshift zero BCGs at redshifts 0.5 (blue), 1 (red), 2
(green), and 3 (orange).  The black histogram shows the stellar mass
distribution of local BCGs. The figure clearly shows that the mass distribution
of progenitors of local BCGs becomes broader and more skewed towards lower
masses with increasing redshift.  At redshift $\sim 1$ the massive tail of
the progenitors of local BCGs are about a factor $3$ smaller than redshift zero
BCGs.  This factor rises to about $10$ at $z\sim 3$.

\section{The evolution of high-L$_{\rm X}$ BCGs}
\label{sec:obs}

In the previous sections, we discussed one facet of the evolution of BCGs,
namely: how did low redshift BCGs form? We now want to address this issue from
an observational perspective.  For example, we want to ask if the low assembly
times measured for our local sample of BCGs are in agreement with observational
results for high-L$_{\rm X}$ clusters. This raises two complementary questions:
\begin{itemize}
\item How do the properties of high-$z$ BCGs differ from those of local ones?
  In other words, {\it what is the observable evolution of BCGs ?}
\item What is the {\it hierarchical relationship} (if any) between BCGs
  observed at different redshifts? In other words, are high-$z$ BCGs
  progenitors of low-$z$ BCGs and are low-$z$ BCGs descendants of high-$z$
  BCGs?
\end{itemize}

We extend the sample used in the previous section by selecting the $125$
central galaxies lying in the most massive haloes at $6$ different redshifts :
$z\sim \{0.2; 0.5; 0.75; 1; 1.5; 2\}$.  Our selection corresponds to a
selection in mass with lower mass limits equal to $M_h / (10^{14} M_\odot) =
\{5.4; 3.7; 2.7; 1.9; 1.2; 0.67\}$ respectively.  We chose this somewhat
arbitrary mass selection so as to preserve the same number density of BCGs at
all redshifts.  This choice would correspond to the (incorrect) monolithic
assumption in which each local BCG has a well-defined dominant progenitor at
each redshift.  Given that we select the most massive haloes at each time-step,
our sample is qualitatively comparable to selecting the highest luminosity
X-ray clusters at each redshift down to a fixed comoving abundance.

\subsection{Observable evolution}
\label{sec:obsevol}

\begin{figure}
\bc
\hspace{-0.64cm}
\resizebox{9cm}{!}{\includegraphics[]{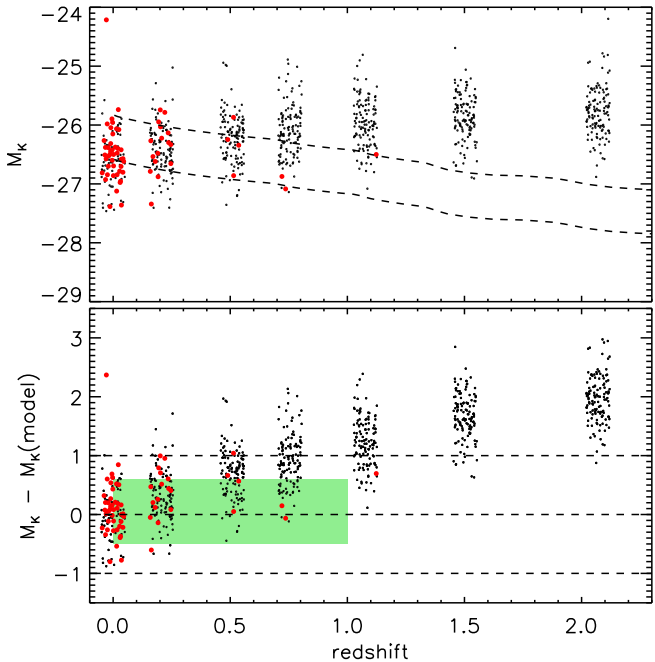}}
\caption{Evolution of the rest-frame $K$-band absolute magnitude of BCGs as a
  function of redshift. Black points show model BCGs (a small random spread has
  been added along the x-axis for clarity), and the red symbols refer to BCGs
  of haloes more massive than $10^{15}\,{\rm M}_{\odot}$.  In the upper panel,
  the dashed lines shows predictions from a single burst model with redshift of
  formation $5$.  The lower line is normalised to the median magnitude of local
  BCGs, and the upper one refers to a normalisation fainter by a factor $2$.
  In the lower panel, black points show the residuals with respect to the model
  with median normalisation.  The shaded (green) area shows the rough location
  of observations (see text for details).}
\label{fig:evol1}
\ec
\end{figure}

In the upper panel of Fig. \ref{fig:evol1}, we show the rest-frame K-band
absolute magnitude distribution of the (low- and) high-$z$ model BCGs. Each
galaxy is represented with a small random offset with respect to its redshift,
for clarity.  The lower dashed line in this panel shows the magnitude decline
predicted by a single burst model where all the stars are formed at $z=5$ and
evolved passively to the present day.  This line is normalised to the median
magnitude of the local sample.  The upper line refers to a normalisation
fainter by a factor $2$.  The model has been built using the
\citet{BruzualCharlot03} code with a solar metallicity stellar population and a
Salpeter IMF with lower and upper mass cutoffs $m_l =
0.1\,{\rm M}_{\sun}$ and $m_u = 100\,{\rm M}_{\sun}$. In the lower panel of
Fig.~\ref{fig:evol1}, we show the residuals of the K-band magnitudes of our
model BCGs with respect to the single burst model (median normalisation), as a
function of redshift.  In both panels, the red symbols highlight BCGs of haloes
more massive than $10^{15}\,{\rm M}_{\odot}$.  Using the local observational
relation between ${\rm M}_{200}$ and L$_{\rm X}$ \citep{Popesso05}, this mass
cut roughly corresponds to the L$_{\rm X} > 2.3\times10^{44}\,{\rm erg}\,{\rm
  s}^{-1}$ cut used by \citet{BurkeCollinsMann00}.

As mentioned before, our low-$z$ BCGs cover the same range of absolute K-band
magnitudes as measured for high-L$_{\rm X}$ clusters.  Model data shown in the
top panel of Fig.~\ref{fig:evol1} indicate that the predicted mass growth
between $z\sim 1$ and $z=0$ is of about a factor $3$.  We note that this is the
same factor measured in Fig.~\ref{fig:stat_ass_form} following the main branch
of local BCGs.  This appears as a coincidence since high-$z$ BCGs are {\it a
  priori} different from progenitors of local BCGs (see next subsection).  This
predicted mass growth also seems to be in conflict with observational results.
The latter are, however, based on high-L$_{\rm X}$ clusters, roughly more
massive than $10^{15}\,{\rm M}_{\odot}$.  Red symbols in the top panel of
Fig.~\ref{fig:evol1} show that selecting haloes above a fixed mass cut
corresponds to selection of ``rarer'' systems at higher redshifts.  For model
BCGs sitting in these haloes, the mass growth is less than a factor $2$ between
$z=1$ and $z=0$.  Although based on low-number statistics, this result is
compatible with that inferred from observational data.

In the bottom panel of Fig.~\ref{fig:evol1}, we reproduce the analysis by
\citet{BurkeCollinsMann00}.  The shaded (green) area in this panel marks the
rough location of their results.  The red symbols again show that a selection
above a fixed mass cut gives a good agreement with these observational data.
The evolution of `rarest' BCGs also appears to be in agreement with passive
evolution.  We find, however, that this does not hold if the selection is
extended to lower mass systems.

The above comparisons should be considered only as a qualitative indication. 
\citet{BurkeCollinsMann00} use observed aperture magnitudes (measured adopting
a $50\,{\rm kpc}$ diameter aperture) and make no attempt to remove the flux due
to other galaxies falling within the aperture.  They also use a different model
normalisation along with a number of different details.  

\begin{figure}
\bc
\hspace{-0.6cm}
\resizebox{9cm}{!}{\includegraphics[]{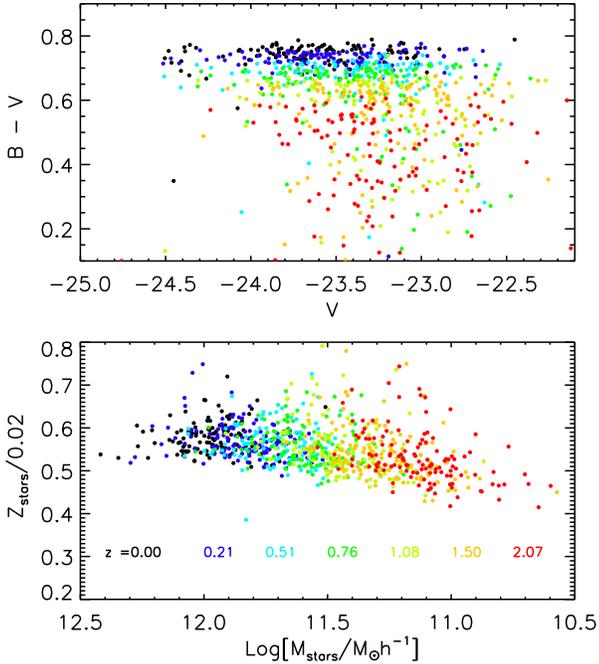}}
\caption{{\it Upper panel} : colour-magnitude relation for BCGs as a
  function of redshift. Symbols are colour-coded with increasing redshift from
  black to red. {\it Lower panel} : mass-metallicity relation, and its
  evolution as a function of redshift.}
\label{fig:evol2}
\ec
\end{figure}

In Fig.~\ref{fig:evol2} we show the rest-frame colour--magnitude (AB) relation
(upper panel) and the relation between the stellar mass and metallicity (lower
panel) for our model BCGs.  Symbols are colour--coded with increasing redshift
from black to red.  The amount of reddening we measure between redshift $\sim
2$ and redshift $0$ is $\sim 0.2$ (considering the median colour for each
redshift bin), which is very close to the amount predicted by the single burst
model used above.  Interestingly the colour--magnitude relation for our model
BCGs has very little slope at any redshift.  The mass--metallicity relation
shown in the bottom panel of Fig.~\ref{fig:evol2} exhibits a more significant,
although still weak, slope.

The results presented in this section change if a different supernovae feedback
model is used.  Our default model in this paper is the supernova feedback model
used in \citet{Croton06}, in which the ejection rate is proportional to the
star formation rate.  We have repeated our analysis using the supernova
feedback model of \citet{DeLucia06}, where the ejection rates are computed on
the basis of energy conservation arguments.  This model results in less
efficient winds in massive galaxies and, as a consequence in more prolonged
star formation activity.  While this does not significantly change results
presented in previous sections, it results in slower evolution of the K-band
magnitudes as a function of redshift which brings more model points into
agreement with the observational data.  However, the K-band magnitudes of local
BCGs are then predicted to be about $0.8$~mag brighter and are typically more
luminous than observed systems.  This different feedback scheme also produces a
larger scatter in the colours and metallicities of model BCGs at all redshifts,
as well as stellar metallicities that are larger than those shown in
Fig.~\ref{fig:evol2} by a factor of $\sim 2$.  The former is a consequence of
the more extended star formation activity, the latter of the processing of
larger amounts of gas into stars.

\subsection{Hierarchical evolution}

In a hierarchical scenario, high-$z$ BCGs are not necessarily the progenitors
of local ones. In this section we want to assess the overlap between these two
populations.

\begin{figure}
  \bc \hspace{-0.6cm}
  \resizebox{9cm}{!}{\includegraphics[]{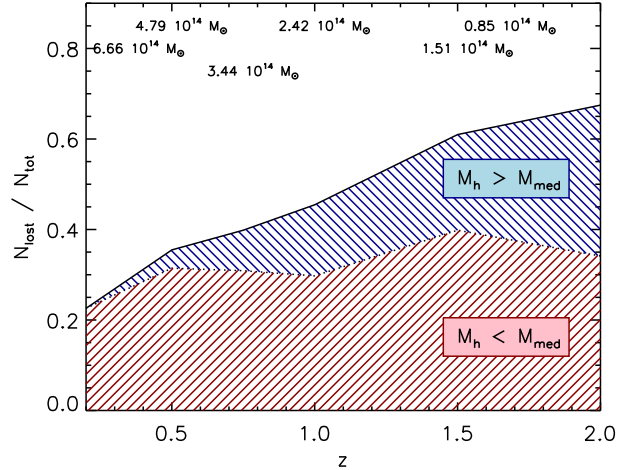}}
\caption{Fraction of high-$z$ BCGs that do not end up in our sample of
  local BCGs. The red area shows contribution from BCGs residing in haloes with
  mass lower than the median mass measured for the sample at each of the
  redshifts considered.  The blue area shows the contribution from massive
  haloes.}
\label{fig:landf}
\ec
\end{figure}

\begin{figure}
\bc
\hspace{-0.6cm}
\resizebox{9cm}{!}{\includegraphics[]{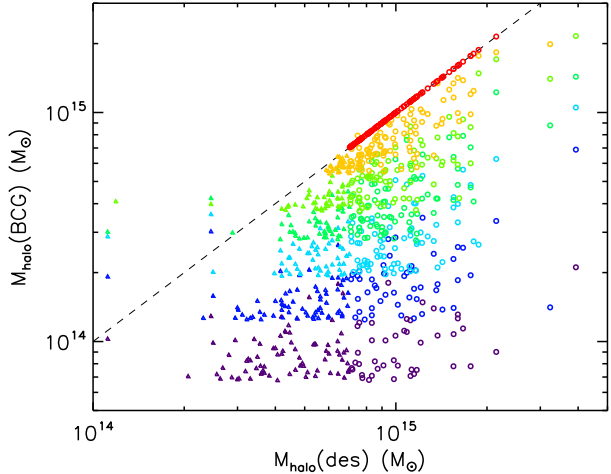}}
\caption{Halo mass distributions. The y-axis shows the masses of
  haloes hosting BCGs at the redshift they are observed; the x-axis shows the
  masses of haloes containing their descendants at $z=0$. Triangles indicate
  ``lost" BCGs and open circles indicate BCGs that end-up in local BCGs. The
  colour-coding indicates redshift, decreasing from purple to red.}
\label{fig:mass}
\ec
\end{figure}

A first question is: what is the fraction of high-$z$ BCGs which {\it do not}
end-up in our local sample? In Fig. \ref{fig:landf}, we show this fraction as a
function of redshift. We have highlighted in red (resp.  blue) the contribution
of BCGs residing in haloes less (resp. more) massive than the median mass of
our samples at each redshift, as indicated on the plot.  A striking remark from
Fig. \ref{fig:landf} is that $\sim 70$ per cent of our BCGs at $z=2$ do not
become part of local BCGs, and this does not appear to depend strongly on the
mass of their host haloes.  However, this fraction of ``lost'' BCGs decreases
with redshift, as expected, and does so more rapidly for BCGs sitting in more
massive haloes.  We note that high-$z$ BCGs, although they do not necessarily
end up in our local BCGs, do end up in massive galaxies at the centre of
massive haloes at $z=0$.  This is illustrated in Fig.  \ref{fig:mass}, which
shows the mass of haloes harbouring high-$z$ BCGs versus the mass of the haloes
harbouring their descendants at $z=0$. In this figure, open triangles indicate
``lost" BCGs, whereas open circles indicate BCGs that end-up in our local
sample.  This figure shows that descendents of high-$z$ BCGs simply extend
towards lower masses our local sample of clusters and we emphasise that this
does not suggest that ``lost" BCGs are in any way a {\it different}
population. 

\begin{figure*}
\bc
\hspace{-0.6cm}
\resizebox{18cm}{!}{\includegraphics[]{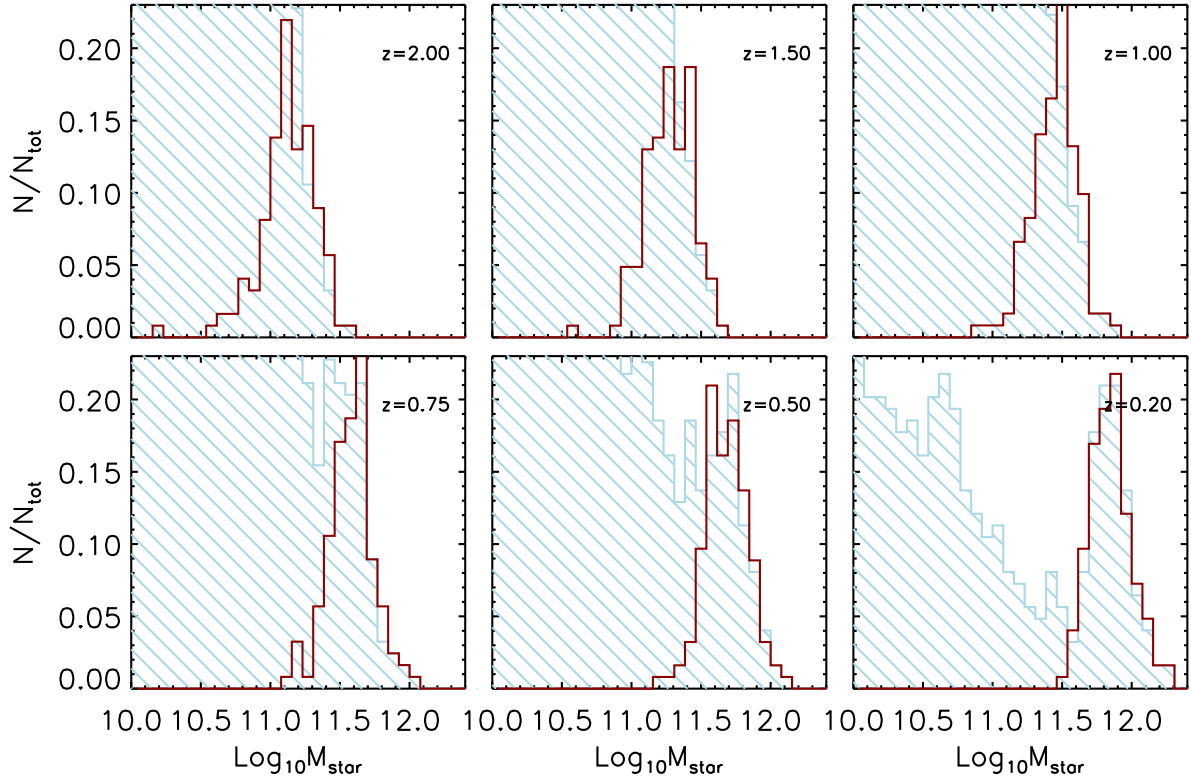}}
\caption{Stellar-mass distribution of high-$z$ BCGs (red histograms) and of
  progenitors of local BCGs (blue histograms). Both histograms are normalised to
  the total number of BCGs at each redshift (125).}
\label{fig:masses}
\ec
\end{figure*}

Turning the question around, we can assess whether the massive progenitors of
our local sample are different from high-$z$ BCGs.  Fig.~\ref{fig:masses} shows
the stellar mass distribution of high-$z$ BCGs (red histograms), and that of
the progenitors of our local BCGs (blue histograms). These distributions are
both normalised to the number of BCGs at each redshift (i.e. 125). Overlap of
the histograms does not imply that they include the same galaxies, but just
that objects of given mass have equal number densities. At redshift lower than
$\sim 0.7$, model BCGs are `assembled' (see Fig.  \ref{fig:stat_times}), and
indeed, the main branch stands out at the massive end of the blue histograms.
There is then a rather good overlap between the mass distributions of high-$z$
BCGs and the massive progenitors of local BCGs.  Moving to higher redshift, the
massive tail of progenitors disappears. As we showed, this is because typically
at $z \gtrsim0.7$ no branch has yet become dominant in the trees of local BCGs.
At these redshifts, high-$z$ BCGs still have a similar mass distribution as the
massive progenitors of local BCGs, although Fig. \ref{fig:landf} shows they are
not the same objects.

At first sight, it may be surprising that our high-$z$ BCGs do not end up in
our local sample. However, we showed that they do belong to the same population
as progenitors of local BCGs, and do end up as central galaxies of massive
haloes. The evolution seen in Fig. \ref{fig:landf} then appears to be
essentially driven by the evolution of dark matter haloes. The most massive
haloes at $z=2$ are not {\it in general} the progenitors of the most massive
haloes at $z=0$ and therefore the BCGs they contain just fall out of our local
sample.

\section{Conclusions and discussion}
\label{sec:concl}

In this study, we have used a combination of high-resolution $N$-body
simulations and semi-analytic techniques to investigate the formation and
evolution of BCGs in a cosmological context. This allows us to explicitly take
into account the full hierarchy of dark matter structure growth, and its impact
on galaxy formation. We show that in a CDM Universe, this hierarchical build-up
implies that central galaxies of massive haloes have complex merging histories
that are not well described by monolithic approximations.

In the first part of our paper, we have studied in detail the formation and
assembly histories of local BCGs. We find that the stars in these systems
start forming very early (about $50$ per cent of the stars in these systems
have already formed by redshift $\sim 5$) although the {\it identity} of the
BCGs themselves is not defined before $z \sim 0.7$.  A monolithic approximation
clearly fails to describe these objects for most of their history.  Most of the
stars are formed in separate entities in a quiescent mode and most of the mass
in the local BCGs is gained through accretion of a relatively low number
(typically $10$) of objects more massive than $10^{10}\,h^{-1}{\rm M}_\odot$. 
Most of these accreted objects, with the exception of those accreted at very
high redshift, have very low gas fractions and star formation rates, quite red
colours and an early-type morphology.  Interestingly, the main branches of BCGs
grow fast enough that very few major mergers occur along them, most of the
accretion events being {\it minor mergers}. 

It is worth mentioning that our quantitative results do depend strongly on the
implemented feedback models, both from AGNs and from supernovae. The AGN
feedback model employed here \citep{Croton06} is extremely efficient in
switching off cooling in relatively massive haloes even at high redshifts. At
the same time, the supernova feedback model we use \citep[also
from][]{Croton06} drives strong winds that blow out all the gas from galaxies
on short timescales. The combination of these two processes causes the early
shutdown of star formation in the progenitors of local BCGs. These two
particular feedback implementations are among the most efficient
proposed so far in the literature. The work presented here leaves room for less
drastic alternatives.

In the second part of our work, we have studied the observable evolution of
BCGs by selecting the $125$ most massive haloes at different redshifts.  From
these samples of high-$z$ BCGs, we measure a mass growth of a factor of $\sim
3$ from $z=1$ to $z=0$. This is {\it by coincidence} very close to the growth
we find following the main branch of local BCGs. These mass growth factors are
higher than those inferred from observational data. However, when we use a
selection that more closely matches that of high-L$_{\rm {X}}$ clusters, we
find milder evolution, consistent with the observations.  Interestingly we find
that our model BCGs exhibit, at all redshifts, a weak correlation between
stellar mass and stellar metallicity and an even weaker correlation between
their magnitudes and their colours.  This changes slightly if a different
supernova feedback model is employed. Our high-$z$ BCGs do end up as central
galaxies of massive haloes at $z=0$ but, because of our cluster sample
selection criterion, many of their descendants are not included in our local
BCG sample.  However, we have shown that there is no substantive physical
difference (1) between local BCGs and the descendents of high-$z$ BCGs, and (2)
between high-$z$ BCGs and the most massive progenitors of local BCGs.


\section*{Acknowledgements}
We thank Volker Springel for his amazing work in the post-processing of the
Millennium Simulation, without which this work would have been impossible. We
also thank Gerard Lemson for setting up the GAVO Millennium database and for
his enthusiastic and patient help with it. We also thank Simon White, Guinevere
Kauffmann, and Alfonso Arag\'on-Salamanca for careful reading of the manuscript
and interesting suggestions, and Stephane Charlot for useful discussions and
suggestions about dust attenuation. GDL thanks the Alexander von Humboldt
Foundation, the Federal Ministry of Education and Research, and the Programme
for Investment in the Future (ZIP) of the German Government for financial
support.

\bsp
\label{lastpage}

\bibliographystyle{mn2e}
\bibliography{deluciablaizot_bcg}

\end{document}